\documentclass[12pt]{article}
\usepackage{graphics}
\usepackage{overpic}
\usepackage{multirow}
\usepackage[absolute,overlay]{textpos}
\usepackage{tikz}

\newcommand{\BR}{{\cal B}}

\newcommand{\fz}{f_0(980)}

\newcommand{\psp}{\psi(2S)}

\newcommand{\jpsi}{J/\psi}

\newcommand{\EE}{e^+e^-}
\newcommand{\MM}{\mu^+\mu^-}

\newcommand{\pp}{\pi^+\pi^-}
\newcommand{\kk}{K^+K^-}

\newcommand{\ppjpsi}{\pi^+\pi^- J/\psi}

\newcommand{\infb}{\rm fb^{-1}}
\newcommand{\inab}{\rm ab^{-1}}
\newcommand{\gev}{\rm GeV}
\newcommand{\mev}{\rm MeV}

\newcommand{\gevcs}{{\gev}/c^2}
\newcommand{\mevcs}{{\mev}/c^2}

\newcommand{\beq}{\begin{equation}}
\newcommand{\eeq}{\end{equation}}
\newcommand{\bitm}{\begin{itemize}}
\newcommand{\eitm}{\end{itemize}}

\def\Journal#1#2#3#4{{#1} {\bf #2} (#4) #3}

\def\PRL{Phys. Rev. Lett.}
\def\PRD{Phys. Rev. D}

\def\EPJC{Eur. Phys. J. C}

\def\CPC{Chin. Phys. C}

\newcommand\pubnumber{WSU--HEP--XXYY}
\newcommand\pubdate{\today}

\def\vt{Department of Physics\\
Virginia Polytechnic Institute and State University\\
Blacksburg, VA 24061-0435, USA}

\def\Title#1{\begin{center} {\Large #1 } \end{center}}
\def\Author#1{\begin{center}{ \sc #1} \end{center}}
\def\Address#1{\begin{center}{ \it #1} \end{center}}

\newcommand\pubblock{\rightline{\begin{tabular}{l} \pubnumber\\
         \pubdate  \end{tabular}}}
\newenvironment{Abstract}{\begin{quotation}  }{\end{quotation}}
\newenvironment{Presented}{\begin{quotation} \begin{center} 
             PRESENTED AT\end{center}\bigskip 
      \begin{center}\begin{large}}{\end{large}\end{center} \end{quotation}}
\def\Acknowledgements{\bigskip  \bigskip \begin{center} \begin{large}
             \bf ACKNOWLEDGEMENTS \end{large}\end{center}}

\def\beq{\begin{equation}}
\def\eeq#1{\label{#1}\end{equation}}
\def\eeqn{\end{equation}}

\def\beqa{\begin{eqnarray}}
\def\eeqa#1{\label{#1}\end{eqnarray}}
\def\eeqan{\end{eqnarray}}

\let\bar=\overbar

\def\Dslash{\not{\hbox{\kern-4pt $D$}}}
\def\dslash{\not{\hbox{\kern-2pt $\del$}}}

\def\BR{\mbox{\rm BR}}

\def\msb{{\bar{\ssstyle M \kern -1pt S}}}

\begin{document}
\begin{titlepage}
\pubblock

\vfill
\Title{Charmonium and exotics from Belle}
\vfill
\Author{Xiaolong Wang}

\Address{\vt}
\vfill
\begin{Abstract}
The recent results on charmonium and charmoniumlike states from Belle experiment are reviewed.
We summarise searches in $B$ decays for possible $X$-like states in final states containing $\eta_c$.
The new measurement of $\EE\to\pp\psp$ improves the determination of the properties of $Y(4360)$
and $Y(4660)$. The fits including $Y(4260)$ are performed. Evidence for a charged charmoniumlike
structure at $4.05~\gevcs$ is observed in the
$\pi^{\pm}\psp$ intermediate state in the $Y(4360)$ decays.
Belle also updates the measurement on $\EE\to\kk\jpsi$ via initial state radiation using the full data sample.
Finally, Belle observed $X(3872)$ in $B\to K\pi + X(\to\pp\jpsi)$. 

\end{Abstract}
\vfill
\begin{Presented}
The 7th International Workshop on Charm Physics (CHARM 2015)\\
Detroit, MI, 18-22 May, 2015
\end{Presented}
\vfill
\end{titlepage}
\def\thefootnote{\fnsymbol{footnote}}
\setcounter{footnote}{0}
%

\section{Introduction}

In the past decade, many structures above open charm 
threshold were discovered and most have 
very exotic properties~\cite{review}. Similarly, comparable states with the $b$-quark,
such as $Y_b$, and the $s$-quark---the $Y(2175)$---were observed. 

The discoveries of the tetra-quark states, $Z^+_b(10610)$ and $Z^+_b(10650)$~\cite{zb}, with quark
content $u\bar{d}b\bar{b}$, and $Z^+_c(3900)$~\cite{zc3900} and $Z^+_c(4020)$~\cite{zc4020}, with quark content
$u\bar{d}c\bar{c}$, extend our knowledge about hadrons and QCD: there are not only 
quark-antiquark mesons or three-quark baryons. It's very possible that the only requirement on a state 
is zero net color. In this case, there should be states with five, six or even more quarks. 

The Belle experiment ran from 1999 to 2010. In this decade, the world 
record on luminosity of $\mathcal{L} = 2.1\times 10^{34}$ was achieved. The data taken by Belle 
experiment are $121~\infb$ on $\Upsilon(5S)$, $711~\infb$ on the $\Upsilon(4S)$ resonance, $3~\infb$ on $\Upsilon(3S)$,
$25~\infb$ on $\Upsilon(2S)$ and $6~\infb$ on $\Upsilon(1S)$. In addition, to study the continuum 
productions, about $100~\infb$ data were recorded off resonance. The full data sample of the Belle 
experiment is $1~\inab$. The huge data sample enhances Belle's ability to study many processes and
states, many of them very rare and statistically limited. Among these are
the charmonium and exotic states. Recently, Belle performed studies on 1) $X$-like states 
decaying to $\eta_c$ modes; 2) an update on $\EE\to\pp\psp$ via initial state radiation (ISR); 3) an update on $\EE\to\kk\jpsi$
via ISR; and 4) $X(3872)$ in $B\to K\pi + \jpsi$.

\section{$X$-like states decaying to $\eta_c$ modes}

The $X(3872)$ was observed by Belle in $B\to K (\jpsi \pp)$~\cite{choi} and is a very narrow resonant state 
with a mass of $M = 3871.69\pm0.17~\mevcs$ and a width of $\Gamma < 1.2~\mev$ at 90\% Conference Level (C.L.)~\cite{pdg}. 
The quantum numbers $J^{PC} = 1^{++}$, favored by Belle, were determined conclusively by LHCb from an angular analysis~\cite{lhcb-x}. 
The nature of the $X(3872)$ remains unclear. One explanation is that $X(3872)$ is a $D^0\bar{D}^{*0}$ molecule;
if so, other ``$X$-like" particles should exist.

Assuming a pair of $D^0$ or $D^{0*}$ can form a $X$-like particle, 
the possible particles are listed in Table~\ref{x-like}.
Belle searched for these possible ``$X$-like" particles in $B^{\pm}\to K^{\pm} X$ decays with 
$X$ decaying to final states containing a $\eta_c$. In the analysis, the test mode $B^{\pm}\to 
K^{\pm}\psp(\to \jpsi\pp)$ was studied first, and the results are quite consistent with 
world-average values~\cite{pdg}.

\begin{table}[t]
\begin{center}
\footnotesize
\caption{The possible $X$-like states and their properties.}
\begin{tabular}{|c | c | c | c|}
\hline
Candidate     & Combination & Quantum number $J^{\rm PC}$ & Decay modes \\\hline
$X_1(3872)$   & $D^0\bar{D}^{*0} - \bar{D}^0D^{*0}$  &  $1^{+-}$  & $\eta_c\omega$, $\eta_c\rho$ \\\hline
$X(3730)$     & $D^0\bar{D}^0 + \bar{D}^0D^0$        &  $0^{++}$  & $\eta_c\eta$, $\eta_c\pi^0$\\\hline
$X(4014)$     & $D^{*0}\bar{D}^{*0} + \bar{D}^{*0}D^{*0}$   & $0^{++}$ & $\eta_c\eta$, $\eta_c\pi^0$\\\hline
\end{tabular}
\label{x-like}
\end{center}
\end{table}

No signal was observed in the final states, however.
The same final states without an $X$-like state were also studied. 
The upper limits of $\BR(B^{\pm}\to K^{\pm}+\eta_c h)$ were determined at 90\% C.L.; 
here, $h$ stands for $\pp$, $\omega$, $\eta$ or $\pi^0$.

\begin{figure}[htb]
\centering
\includegraphics[width=13.0cm]{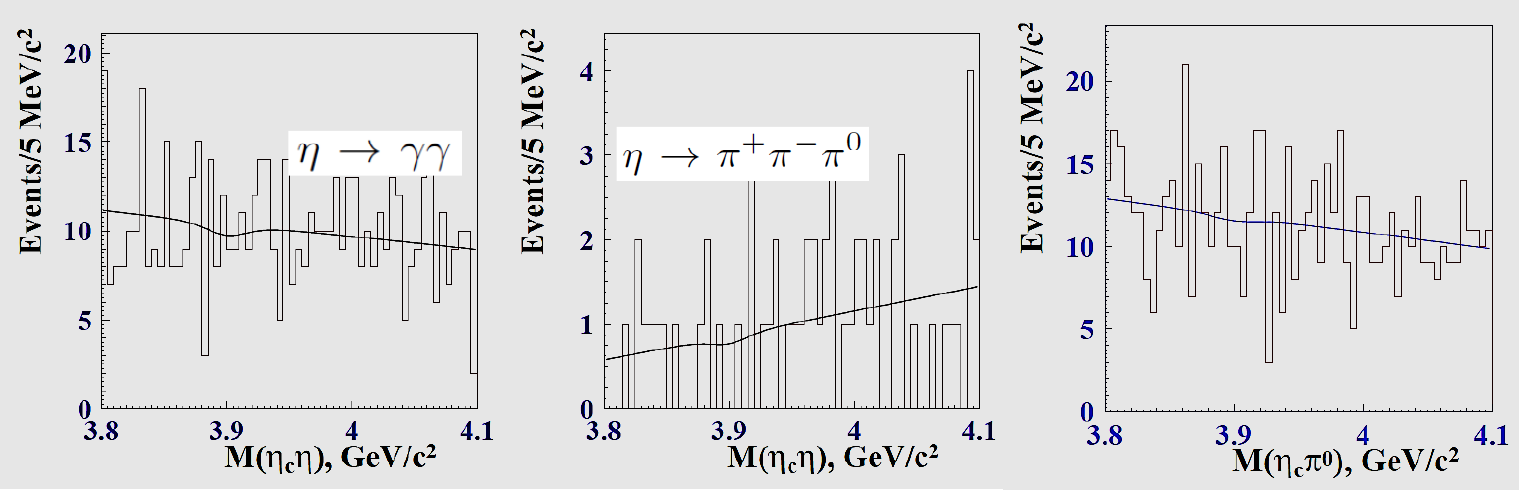}
\caption{The distributions of $M_{\eta_c \eta}$ and $M_{\eta_c\pi^0}$ in the decays of 
$B^{\pm}\to K^{\pm}+X$ at Belle. The curves show the fits with a possible $Z_c^0(3900)$.}
\label{etac-2}
\end{figure}

Since $Z^{\pm}_c(3900)$ was observed in $\pi^{\pm}\jpsi$ final states~\cite{zc3900} 
and $Z^{\pm}_c(4020)$ was observed in $\pi^{\pm}h_c$ 
final states~\cite{zc4020}, they may have neutral partners analagous to the observed
$Z^0_b(10610)$ and $Z^0_b(10650)$.
They may decay to final states containing an $\eta_c$ meson. Additionally, this search is
sensitive to the $X(3915)$ that was discovered in 
$\gamma\gamma$ collisions~\cite{x3915}. Belle doesn't observe signal in the final 
states, as shown in Fig.~\ref{etac-2}.

\section{Update on $\EE\to\pp\psp$ via ISR}

$Y(4360)$ was confirmed and $Y(4660)$ was discovered in the Belle measurement 
of $\EE\to\pp\psp$ via ISR~\cite{pppsp-belle1}. $Y(4660)$ has been confirmed recently
by BaBar~\cite{pppsp-babar2}. 
$Y(4660)$ is the highest-mass charmoniumlike state; oddly, its width is only $(48\pm15\pm9)~\mev$,
according to Belle's previous measurement~\cite{pppsp-belle1}. Additionally, the structure $X(4630)$ observed in 
$\EE\to\Lambda_c\bar{\Lambda}_c$~\cite{y4630} has a mass close to that of $Y(4660)$; it remains
unclear whether they are the same resonance.
With an improved reconstruction efficiency, Belle's full data set yields additional sensitivity
to this question in its to update the measurement of 
$\EE\to\pp\psp$. Meanwhile, the $Z^+_c(3900)$ observed in $Y(4260)$ decays is a good candidate of 
tetra-quark state~\cite{zc3900}. Since $Y(4360)$ and $Y(4660)$ share some similar properties with $Y(4260)$, 
it is important to search for this pattern of possible intermediate state(s) in $Y(4360)$ or $Y(4660)$ decays.

\begin{figure}[htb]
\centering
\includegraphics[width=6.0cm]{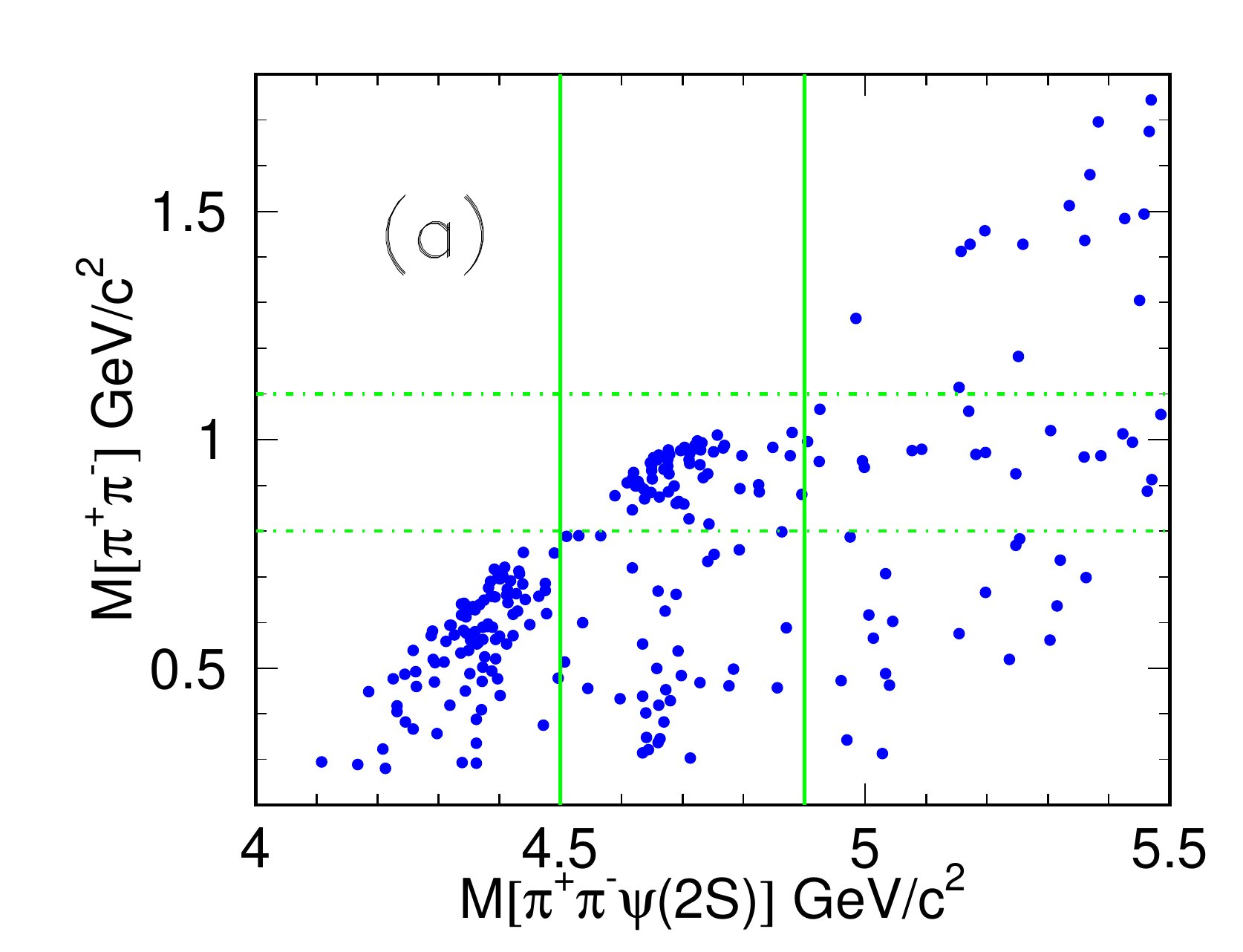}
\includegraphics[width=6.0cm]{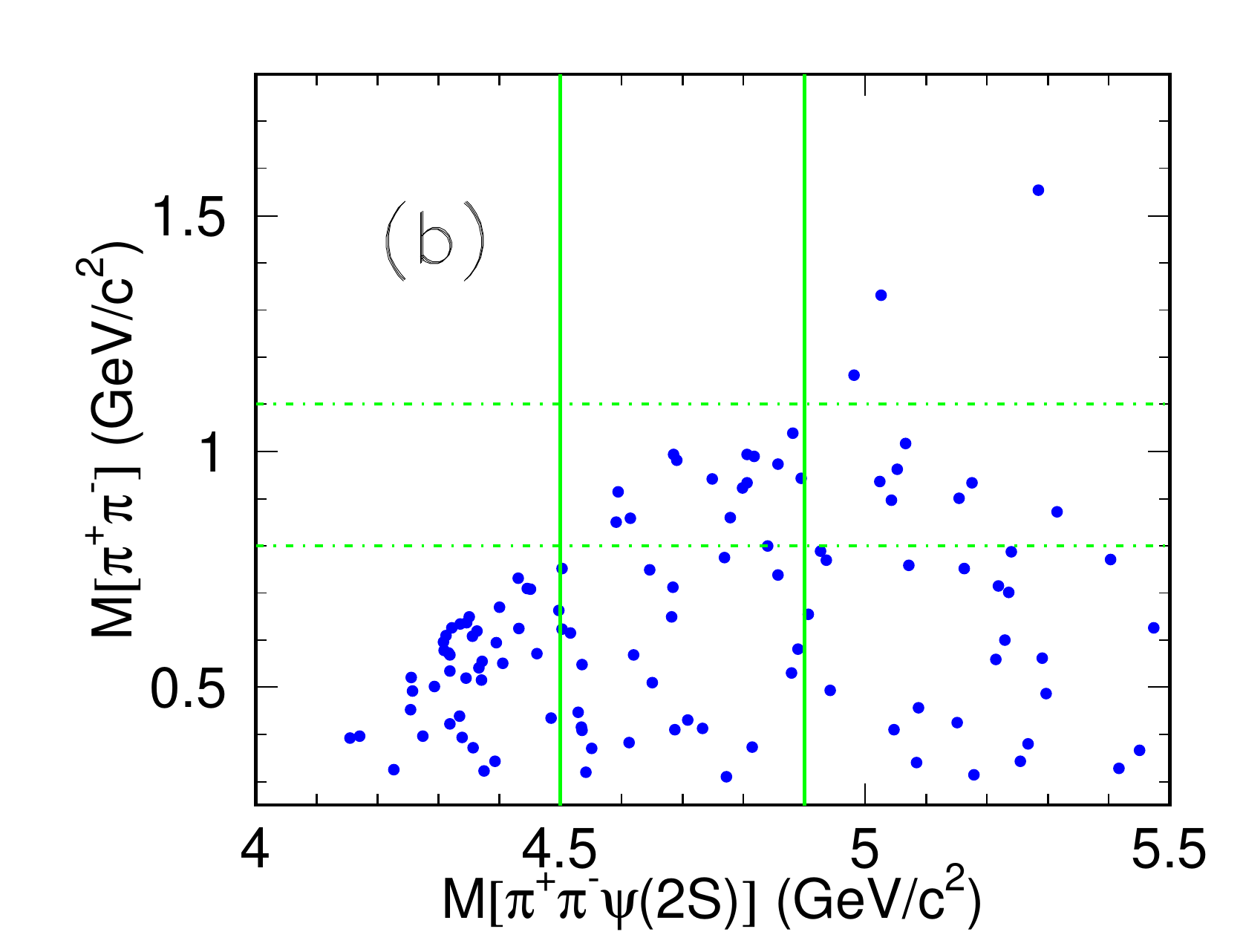}
\caption{Invariant mass of the $\pp$ recoiling against the $\psp$
 versus the invariant mass of the $\pp\psp$
 in the $\ppjpsi$ mode (a) and $\MM$ mode
(b). The horizontal dashed lines show the belt of $\fz$,
while the vertical solid lines demarcate the regions
with the $Y(4360)$ and $Y(4660)$ states and
the higher-mass combinations.}
\label{pppsp-scat}
\end{figure}

\begin{figure}[htb]
\centering
\includegraphics[width=3.5cm,angle=-90]{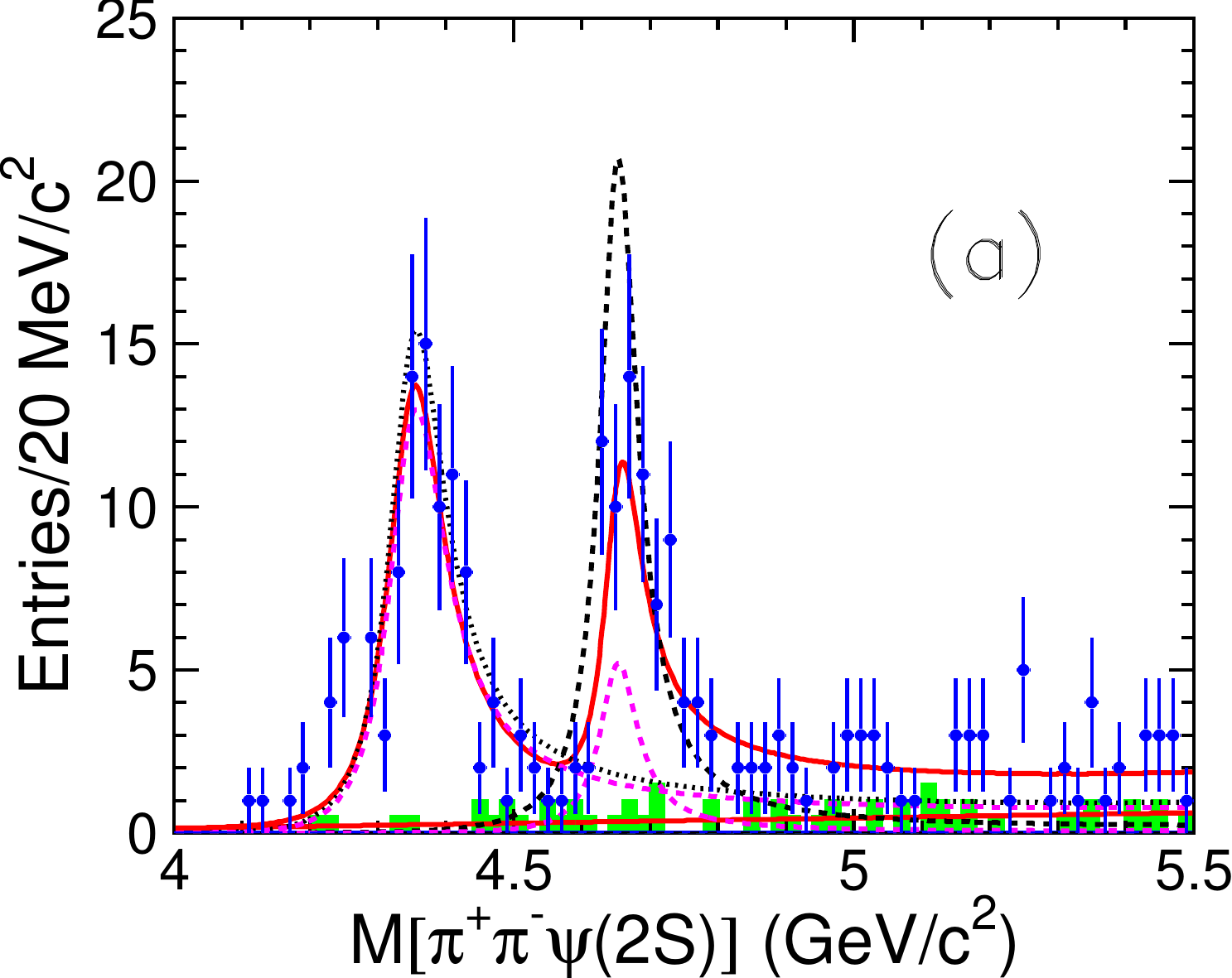}
\includegraphics[width=3.5cm,angle=-90]{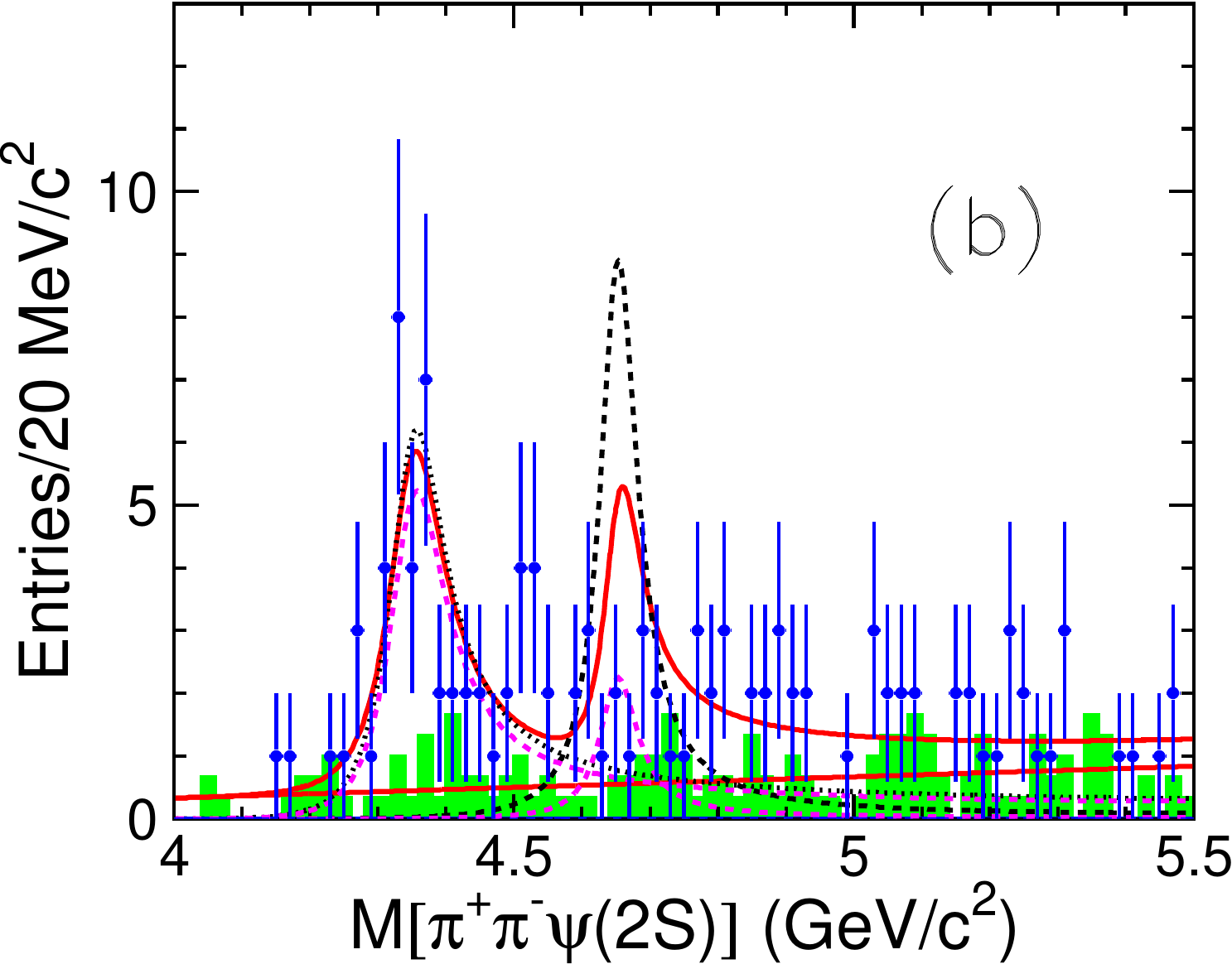}
\includegraphics[width=3.5cm,angle=-90]{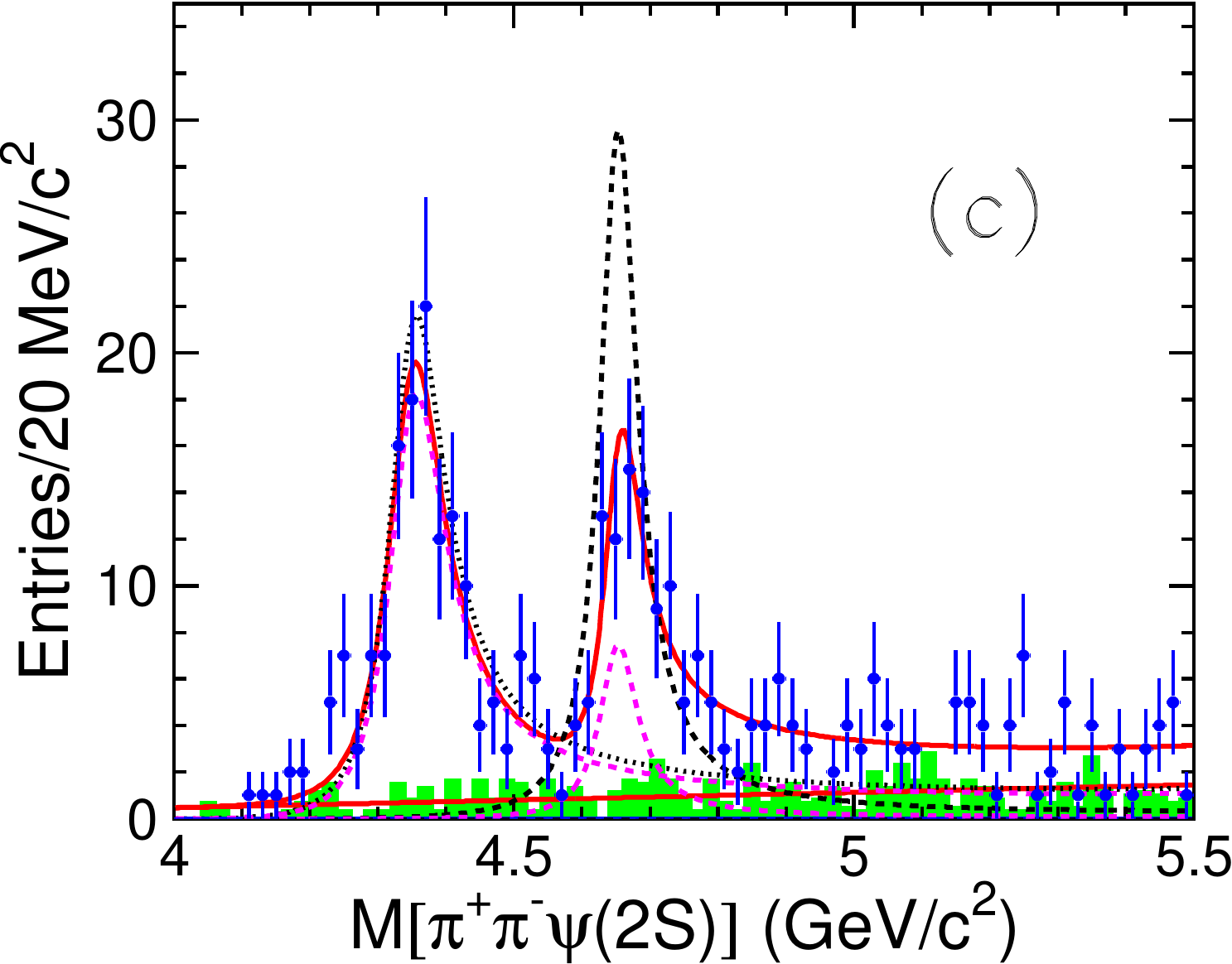}
\caption{The $\pp\psp$ invariant-mass distributions and the
simultaneous fit results. From left to
right: (a) the $\ppjpsi$ mode, (b) the $\MM$ mode,
and (c) the sum.}
\label{pppsp-fit-plot}
\end{figure}

Belle uses the full data sample to update the measurement on $\EE\to\pp\psp$~\cite{pppsp-belle2}. 
$\psp\to\pp\jpsi$ ($\ppjpsi$ mode) and $\psp\to\MM$ ($\MM$ mode) are used to reconstruct the $\psp$ signal. 
Belle sees 245 candidate events with a purity of 96\% from the $\ppjpsi$ mode
and 118 events with a purity of 60\% from the $\MM$ mode. Figure~\ref{pppsp-scat} shows the scatter plots of 
$M_{\pp\psp}$ vs. $M_{\pp}$.
Unbinned simultaneous maximum likelihood fit are performed to $M_{\pp\psp}$ for $Y(4360)$ and $Y(4660)$ using 
$Amp = BW_1+e^{i\phi}\cdot BW_2$. The fit results are shown in Fig.~\ref{pppsp-fit-plot} 
and Table~\ref{pppsp-fit-tab}.
They are consistent with the previous measurement~\cite{pppsp-belle1}. No obvious signal is seen
above $Y(4660)$ but some events accumulate at $Y(4260)$, especially in the $\ppjpsi$ mode.

\begin{table}
\caption{Results of the fits to the $\pp\psp$ invariant-mass
spectra. The first error is statistical and the second is
systematic. $M$, $\Gamma$, and $\BR\cdot \Gamma^{\EE}$ are the
mass (in $\mevcs$), total width (in $\mev$), and the product of the
branching fraction to $\pp\psp$ and the $\EE$ partial width (in
eV), respectively; $\phi$ is the relative phase between the two
resonances (in degrees).}\label{pppsp-fit-tab}
\begin{center}
\begin{tabular}{c |c c }
\hline\hline
     Parameters     & ~~~Solution~I~~~ & ~~~Solution~II~~~  \\\hline
 $M_{Y(4360)}$        & \multicolumn{2}{c}{$4347\pm6\pm3$} \\
 $\Gamma_{Y(4360)}$   & \multicolumn{2}{c}{$103\pm9\pm5$}  \\
 $\BR[Y(4360)\to\pp\psp]\cdot\Gamma_{Y(4360)}^{\EE}$
                    & ~~$9.2\pm0.6\pm0.6$~~ & ~~$10.9\pm0.6\pm0.7$~~   \\
 $M_{Y(4660)}$       & \multicolumn{2}{c}{$4652\pm10\pm11$} \\
 $\Gamma_{Y(4660)}$  & \multicolumn{2}{c}{$68\pm11\pm5$}   \\
 $\BR[Y(4660)\to\pp\psp]\cdot\Gamma_{Y(4660)}^{\EE}$
                    & ~~$2.0\pm0.3\pm0.2$~~ & ~~$8.1\pm1.1\pm1.0$~~  \\
 $\phi$             & ~~$32\pm18\pm20$~~ & ~~$272\pm8\pm7$~~  \\
 \hline\hline
\end{tabular}
\end{center}
\end{table}

\begin{figure}[htb]
\centering
\includegraphics[width=3.7cm,angle=-90]{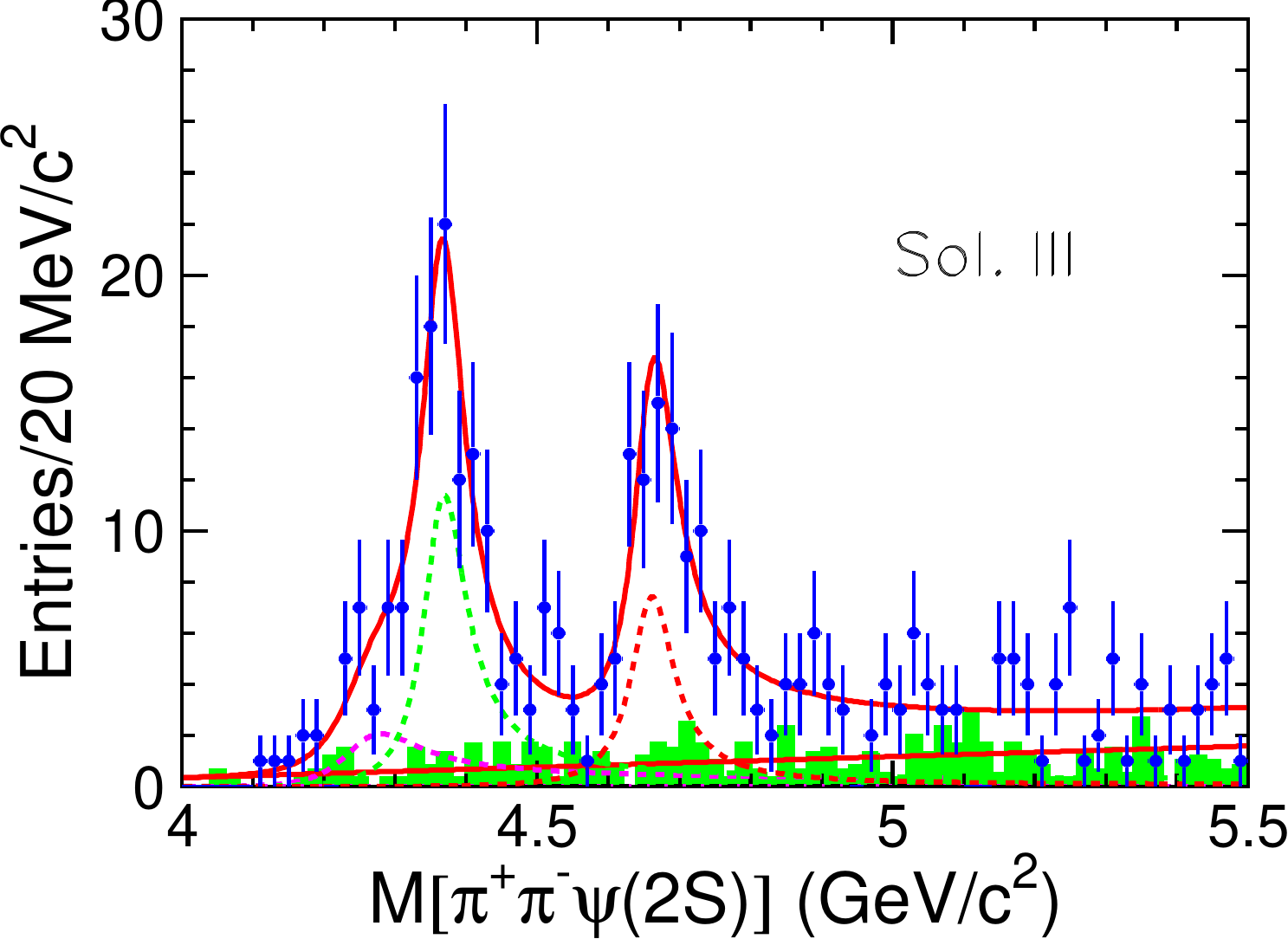}
\includegraphics[width=3.7cm,angle=-90]{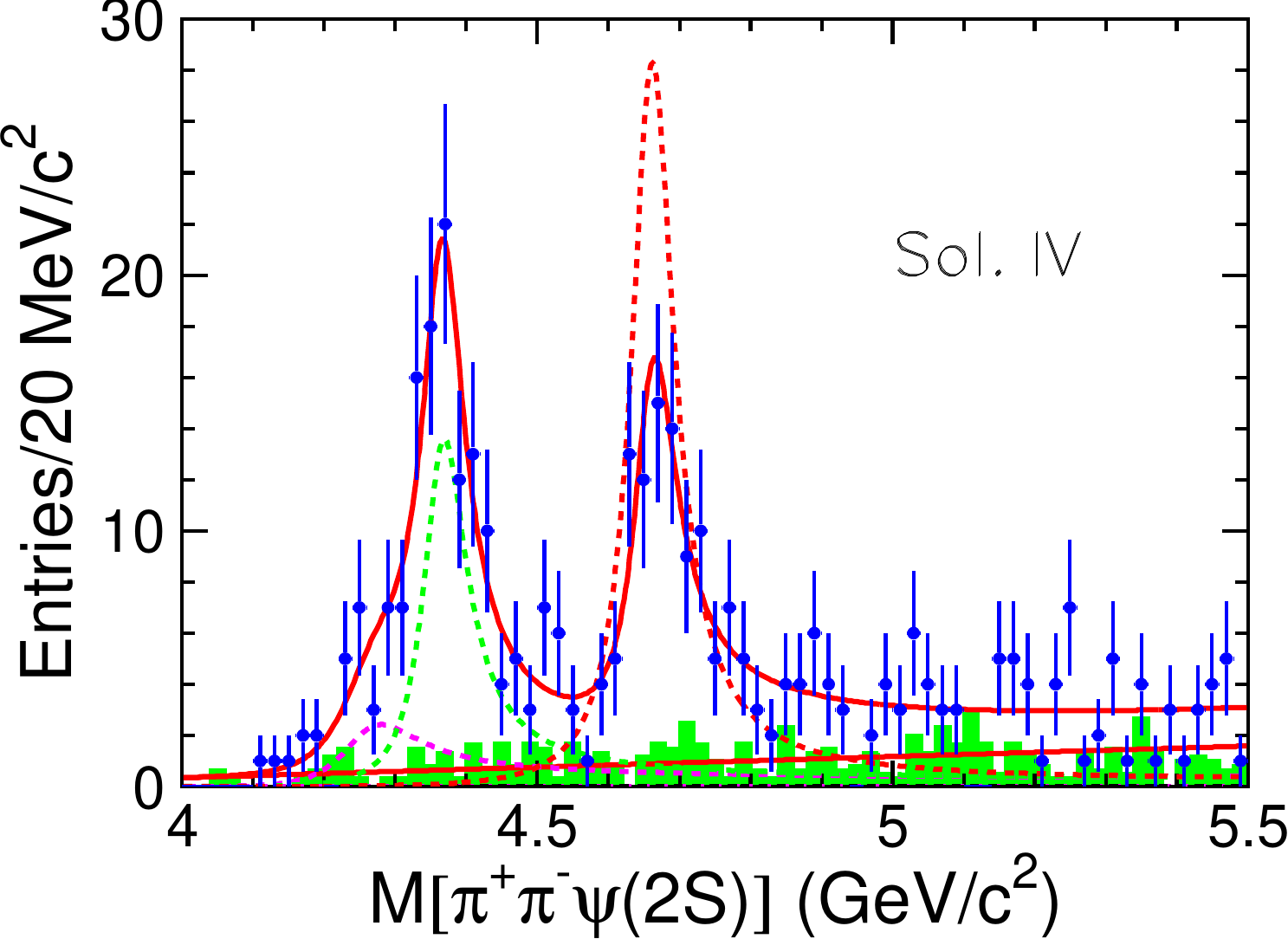}\\
\includegraphics[width=3.7cm,angle=-90]{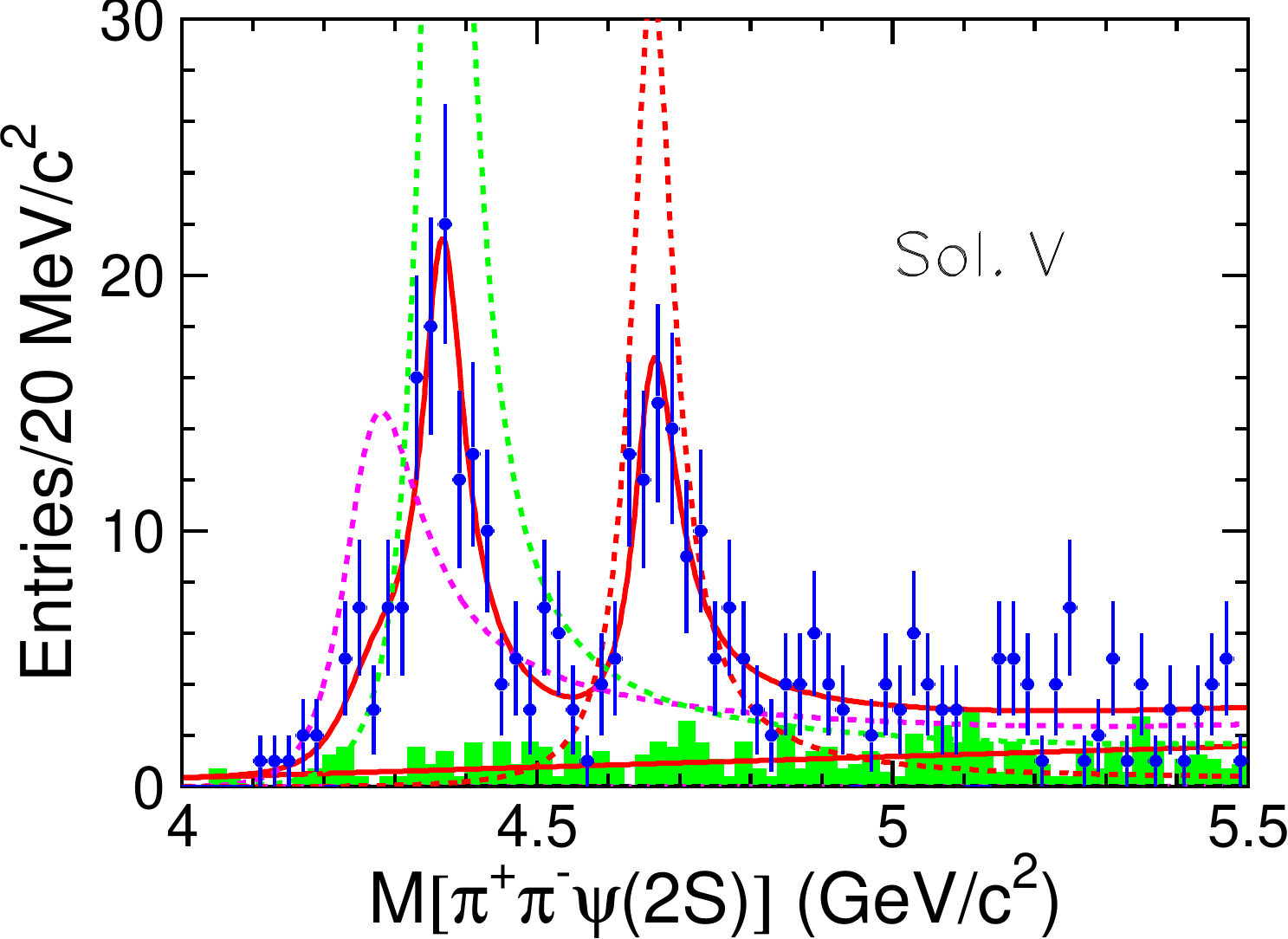}
\includegraphics[width=3.7cm,angle=-90]{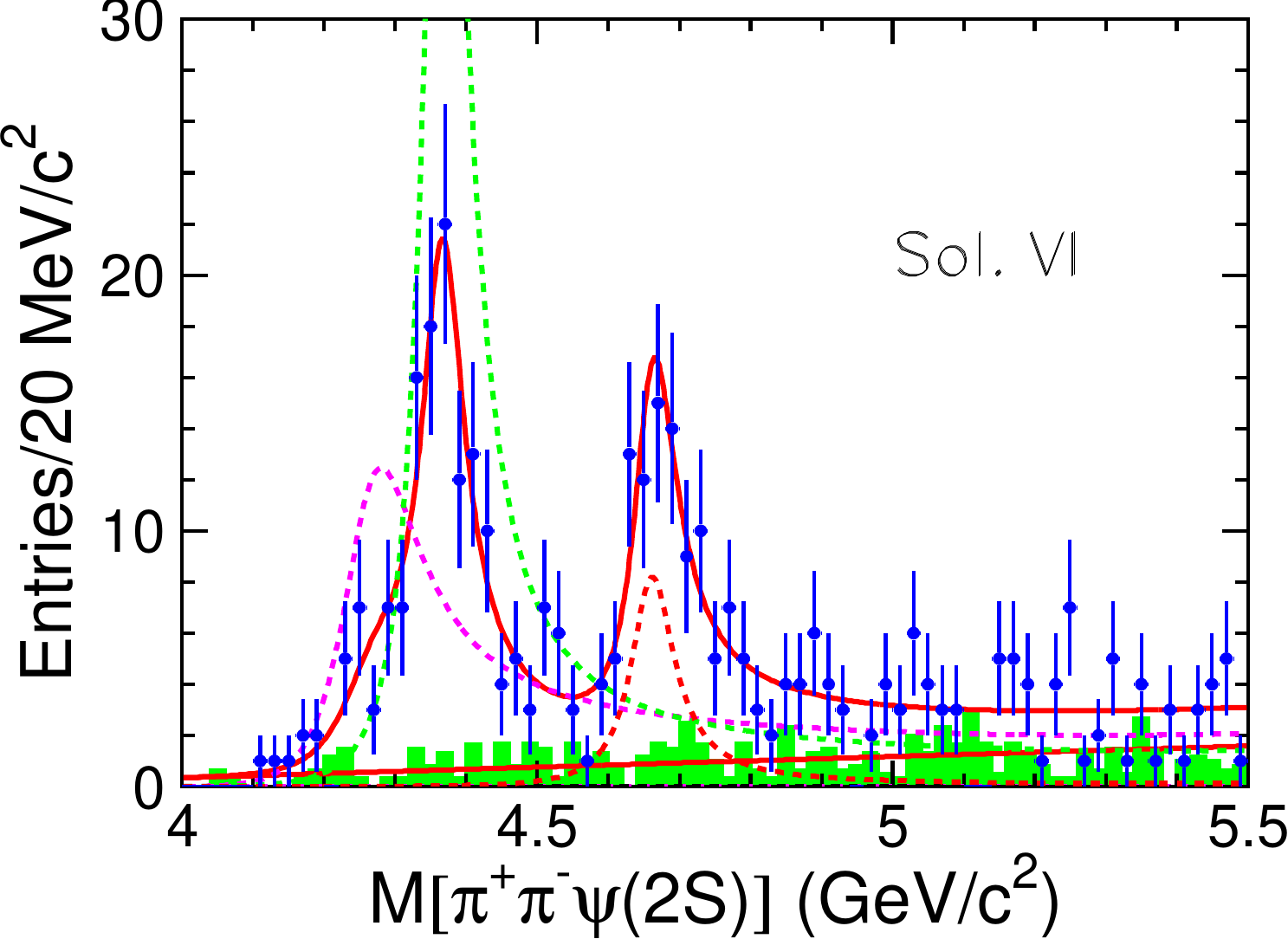}
\caption{The four solutions from the fit to the $\pp\psp$
invariant mass spectra with the $Y(4260)$ included. The curves
show the best fit and the dashed curves show the contributions
from the two Breit-Wigner (BW) components.}
\label{3r-fit-plot}
\end{figure}

Since there are events in the vicinity of the
$Y(4260)$ mass, an alternative fit with a coherent
sum of $Y(4260)$, $Y(4360)$, and $Y(4660)$ amplitudes is performed.
The fit results are shown in Fig.~\ref{3r-fit-plot} and
Table~\ref{three_res}.
The signal significance of the $Y(4260)$ is only $2.4\sigma$ but it affects the masses and widths  
of $Y(4360)$ and $Y(4660)$.

\begin{table*}
\caption{Results of the alternative fits to the $\pp\psp$ invariant-mass
spectra using three resonances: $Y(4260)$, $Y(4360)$, and
$Y(4660)$. The parameters are the same as in Table~\ref{pppsp-fit-tab},
except that, here, $\phi_1$ is the relative phase between the $Y(4360)$
and $Y(4260)$ (in degrees) and $\phi_2$ is the relative phase
between the $Y(4360)$ and $Y(4660)$ (in degrees).}
\label{three_res}
\begin{center}
\begin{tabular}{c |c c c c }
\hline\hline
     Parameters     & ~~Solution~III~~~ & ~~~Solution~IV~~ & ~~Solution~V~~ & ~~~Solution~VI~~ \\\hline
 $M_{Y(4260)}$        & \multicolumn{4}{c}{4259~(fixed)} \\
 $\Gamma_{Y(4260)}$   & \multicolumn{4}{c}{134~(fixed)}  \\
 $\BR\cdot\Gamma_{Y(4260)}^{\EE}$
                    & $1.5\pm0.6\pm0.4$ & $1.7\pm0.7\pm0.5$ & $10.4\pm 1.3\pm0.8$ & $8.9\pm 1.2\pm0.8$  \\
 $M_{Y(4360)}$        & \multicolumn{4}{c}{$4365\pm7\pm4$} \\
 $\Gamma_{Y(4360)}$   & \multicolumn{4}{c}{$74\pm14\pm4$}  \\
 $\BR\cdot\Gamma_{Y(4360)}^{\EE}$
                    & $4.1\pm1.0\pm0.6$ & $4.9\pm1.3\pm0.6$ & $21.1\pm 3.5\pm1.4$ & $17.7\pm 2.6\pm1.5$ \\
 $M_{Y(4660)}$       & \multicolumn{4}{c}{$4660\pm9\pm12$} \\
 $\Gamma_{Y(4660)}$  & \multicolumn{4}{c}{$74\pm12\pm4$}   \\
 $\BR\cdot\Gamma_{Y(4660)}^{\EE}$
                    & $2.2\pm0.4\pm0.2$ & $8.4\pm0.9\pm0.9$ & $9.3\pm 1.2\pm1.0$ & $2.4\pm 0.5\pm0.3$\\
 $\phi_1$             & $304\pm24\pm21$ & $294\pm25\pm23$ & $130 \pm 4\pm2$ & $141\pm 5\pm4$\\
 $\phi_2$             & $26\pm19\pm10$ & $238\pm14\pm21$ & $329\pm 8\pm5$ & $117\pm 23\pm25$\\
 \hline\hline
\end{tabular}
\end{center}
\end{table*}

\begin{figure}[htb]
\centering
\includegraphics[width=5.5cm,angle=-90]{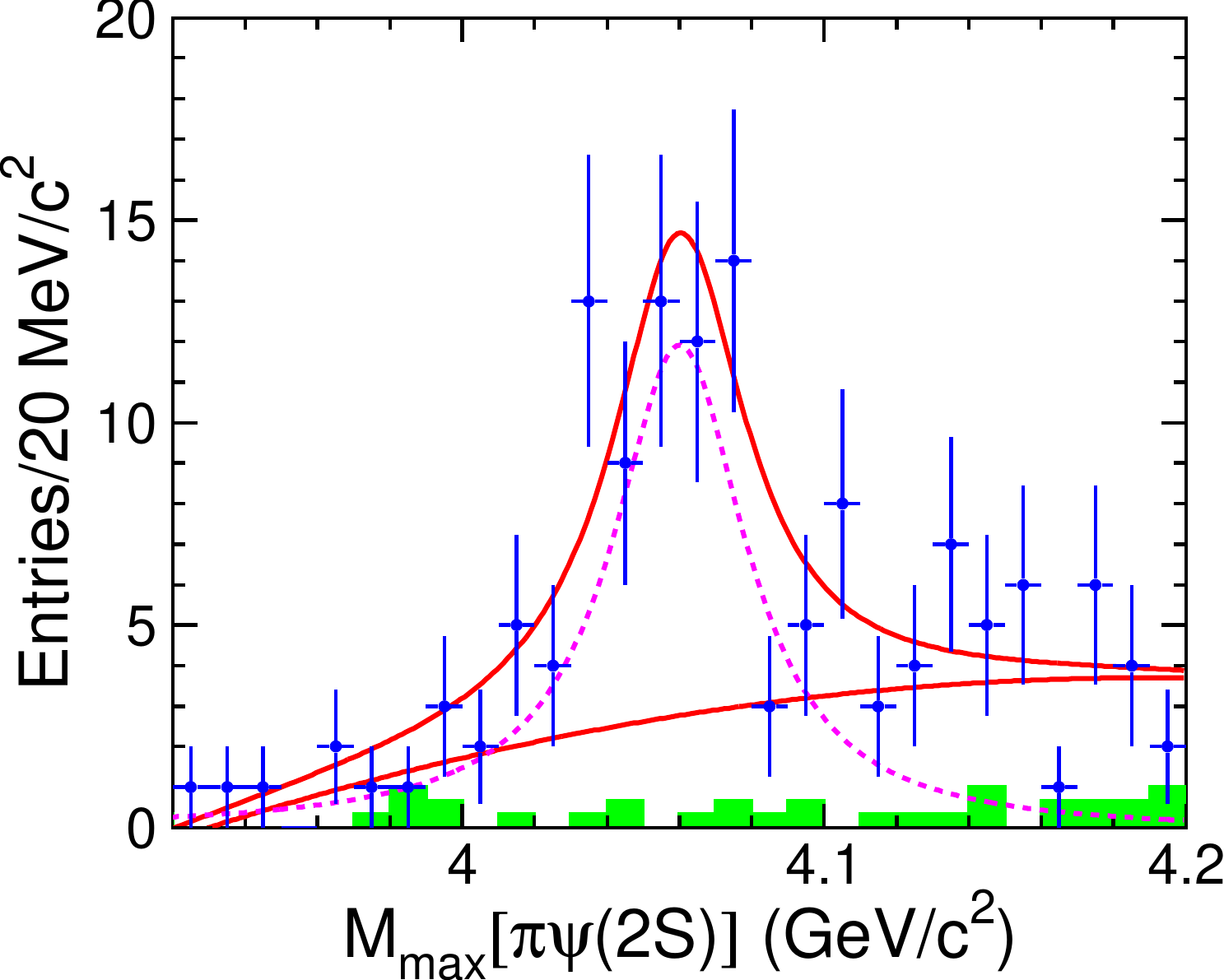}
\caption{The distribution of $M_{\rm max}(\pi^{\pm}\psp)$ from
$Y(4360)$-subsample decays. The solid curve is the
best fit and the dashed curve is the signal parametrized by a
BW function. }
\label{mppsp-fit}
\end{figure}

Belle searches for charged charmoniumlike structures in
both $\psi(2S)$ decay modes of the $\pi^{\pm}\psp$ system
from $Y(4360)$ or $Y(4660)$ decays. In $Y(4360)$ decays, there is an excess evident
at around $4.05~\gevcs$ in the $\pi^\pm\psp$ invariant-mass
distributions in both $\ppjpsi$ and $\MM$ modes. An unbinned maximum-likelihood fit is
performed simultaneously for both modes on the distribution of $M_{\rm max}(\pi^{\pm}\psp)$, the
maximum of $M(\pi^+\psp)$ and $M(\pi^-\psp)$.
The fit yields a mass of $(4054\pm 3({\rm stat.})\pm 1({\rm
syst.}))~\mevcs$ and a width of $ (45\pm 11({\rm stat.})\pm 6({\rm syst.}))~\mev$ for 
the excess, as shown in Fig.~\ref{mppsp-fit}. The significance of this excess is $3.5\sigma$.
No charged structure is found in $Y(4660)$ decays.

\section{Update on $\EE\to\kk\jpsi$ via ISR}

The first scan on this channel was performed by Belle~\cite{kkjpsi}. The cross section of $\EE\to\kk\jpsi$ was measured 
from threshold to $5.5~\gevcs$. Unlike the final states of $\pp\jpsi$ and $\pp\psp$, in which clear 
resonant signals were observed, the distribution of $M_{\kk\jpsi}$ lacked evidence for structure.

\begin{figure}[htb]
\centering
\includegraphics[width=12cm]{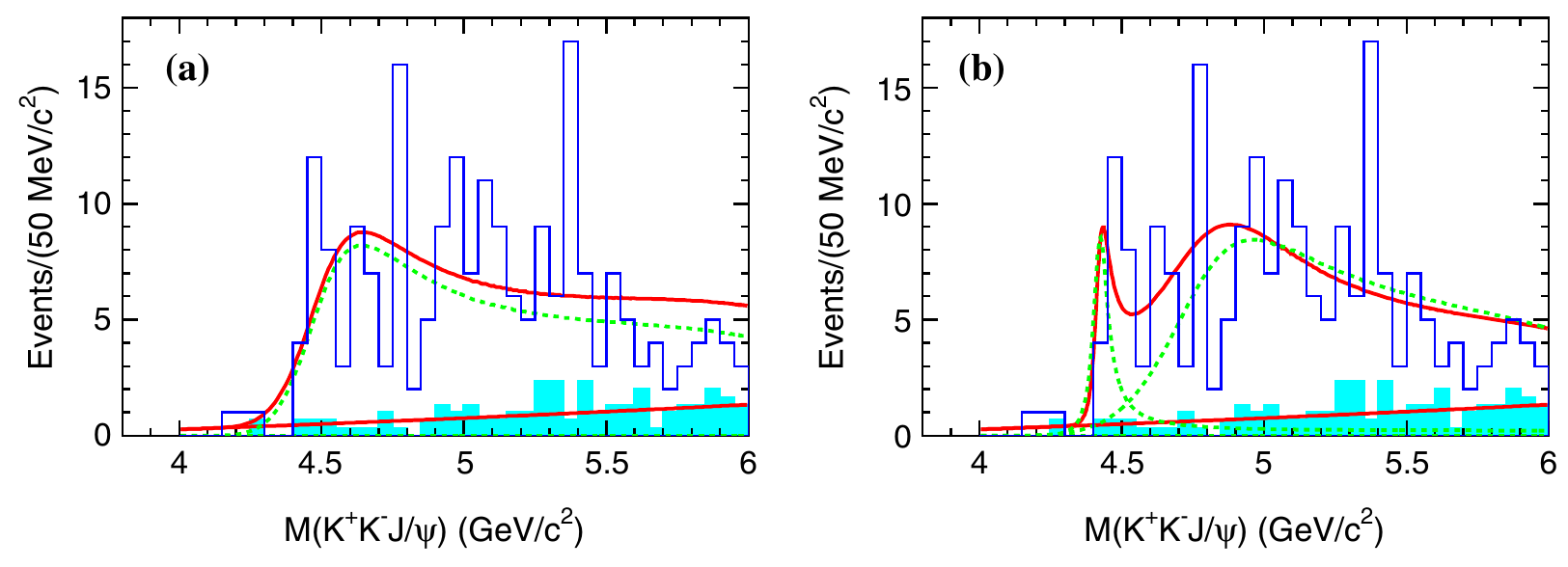}
\caption{Fits to $M_{\kk\jpsi}$ distribution at Belle. 
The solid curves show the best fit to data and background with 
one BW function (a) and the coherent sum of a BW function and the $\psi(4415)$ component (b).}
\label{kkjpsi-fit}
\end{figure}

\begin{figure}[htb]
\centering
\includegraphics[width=15cm]{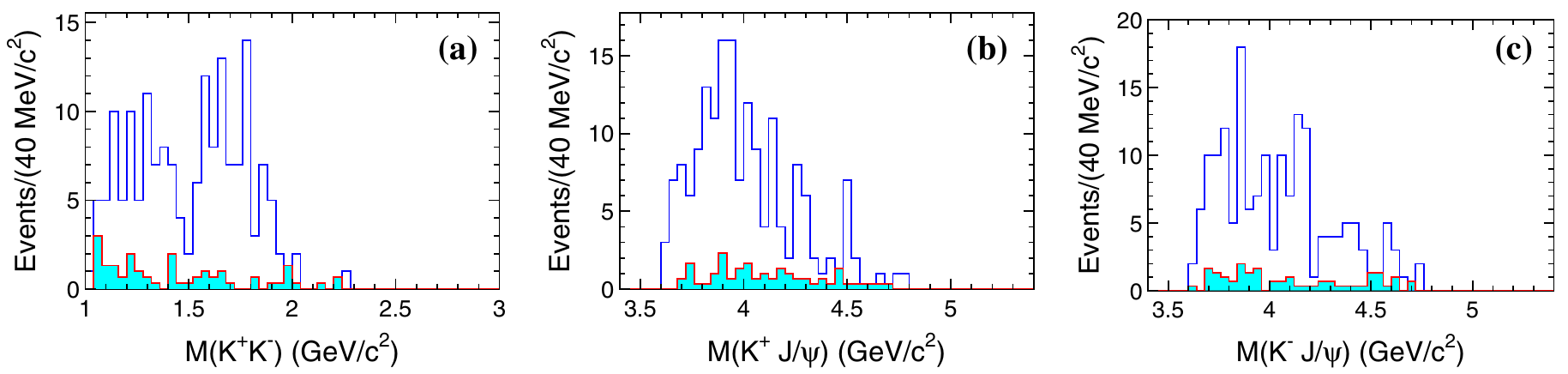}
\caption{The invariant-mass distributions of (a) $\kk$, (b) $K^+\jpsi$ and (c) $K^-\jpsi$.}
\label{kkjpsi-proj}
\end{figure}

Recently, Belle has updated this measurement with the full data sample~\cite{kkjpsi2}. Despite
the enhanced statistics,
there is still no $Y$ state observed in the final states. The $M_{\kk\jpsi}$ distribution is shown in Fig.~\ref{kkjpsi-fit}.
Additionally, Belle searched for $Z_{cs}\to K\jpsi$ in this analysis. As shown in Fig.~\ref{kkjpsi-proj},
no evident structure in $K^{\pm}\jpsi$  mass distribution is observed.

\section{Study of $B\to X(3872)K\pi$}

\begin{figure}[htb]
\centering
\begin{overpic}[width=6cm]{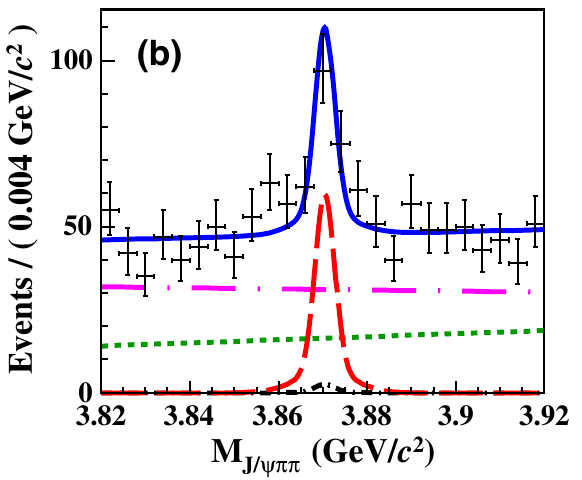}
\unitlength=1mm
\put(13,41){\includegraphics[width=0.6cm]{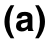}}
\end{overpic}
\begin{overpic}[width=6cm]{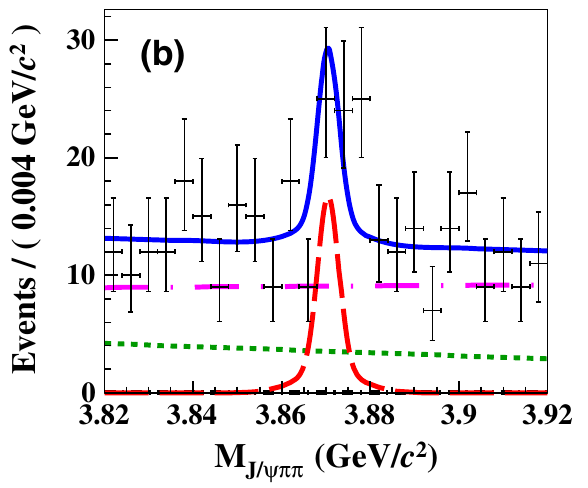}
\put(22,70){\includegraphics[width=0.6cm]{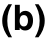}}
\end{overpic}
\caption{$X(3872)$ in $B^0\to X(3872)(\to\ppjpsi)K^+\pi^-$ decays (a) and 
$B^{\pm}\to X(3872)(\to\ppjpsi)K_s\pi^{\pm}$ (b). }
\label{xkpi-belle}
\end{figure}

The $X(3872)$ was discovered in $B^+\to X(3872)(\to\ppjpsi)K^+$~\cite{choi} and has a definitive 
$J^{PC}$ assignment of $1^{++}$~\cite{lhcb-x}. It has been observed to decay to several other final 
states: $\jpsi\gamma$, $\psp\gamma$, $\jpsi\pp\pi^0$ and $D^0\bar{D}^{*0}$. Recently, Belle searched for 
$X(3872)$ in another final state: $B\to X(3872)(\to\ppjpsi)K\pi$~\cite{xkpi-belle}. Figure~\ref{xkpi-belle} shows 
the signals seen in both $B^+$ and $B^0$ decays. The extracted branching fraction is
$\BR(B\to X(3872)K\pi)\times\BR(X(3872)\to\jpsi\pp) =
(7.9\pm 1.3\pm 0.4)\times 10^{-6}$ in $B^0$ decay and $(10.6\pm 3.0\pm 0.9)\times 10^{-6}$
in $B^+$ decay, which agree with each other quite well. In addition, Belle finds that $B^0\to X(3872)K^*(892)^0$ 
does not dominate the $X(3872)K^+\pi^-$ final state.

\Acknowledgements
We would like to thank all of the organizers and participants of CHARM 2015 in making this meeting a success
and acknowledge the support of the organizers in presenting these results.


\end{document}